\documentclass[sigconf]{acmart} 
\usepackage{graphicx}
\usepackage{amsmath}
\usepackage{subcaption}
\usepackage{epstopdf}
\usepackage{amsthm}
\usepackage{booktabs}
\usepackage{hyperref}
\usepackage{multirow}
\usepackage{romannum}
\usepackage{tabularx}

\usepackage{algorithm,algpseudocode}
 
\usepackage{amsfonts}

\setcopyright{none}

\acmDOI{00.000/000_0}
 
\acmISBN{000-0000-00-000/00/00}
 
\acmConference[0000]{000}{0000}{Beijing, China} 
\acmYear{0000}
\copyrightyear{0000}
\acmPrice{00.00}

\begin{document}
	\title{An efficient implementation for solving the all pairs minimax path problem in an undirected dense graph}
	
	\author{Gangli Liu}
	\affiliation{%
		\institution{Tsinghua University}
	}
	\email{gl-liu13@mails.tsinghua.edu.cn}

\begin{abstract}
We provide an efficient $ O(n^2) $ implementation for solving the all pairs minimax path problem or  widest path problem in an undirected dense graph. It is a code implementation of the Algorithm 4 (MMJ distance by Calculation and Copy) in a previous paper. The distance matrix is also called the all points path distance (APPD). We conducted experiments to test the implementation and algorithm, compared it with several other algorithms for solving the APPD matrix.  Result shows Algorithm 4 works good for solving the widest path or minimax path APPD matrix.  It can drastically improve the efficiency for computing the APPD matrix.  There are several theoretical outcomes which claim the APPD matrix can be solved accurately in $ O(n^2) $ . However, they are impractical because there is no code implementation of these algorithms. It seems Algorithm 4 is the first algorithm that has an actual code implementation for solving the APPD matrix of minimax path or widest path problem in $ O(n^2) $, in an undirected dense graph.

\end{abstract}
 
	\maketitle
 
\section{Introduction} \label{sec_one}
The minimax path problem is a classic problem in graph theory and optimization. It involves finding a path between two nodes in a weighted graph such that the maximum weight of the edges in the path is minimized. \footnote{\url{https://en.wikipedia.org/wiki/Widest_path_problem}}

Given a graph \( G = (V, E) \) where \( V \) is the set of vertices and \( E \) is the set of edges, each edge \( e \in E \) has a weight \( e_w \).   For an undirected graph with \(n\) vertices, the maximum number of edges is \(\frac{n(n-1)}{2}\). A dense graph has close to \(\frac{n(n-1)}{2}\) edges. We can say a dense graph has \( O(n^2) \) edges.  In an undirected graph, each edge is bidirectional, meaning it connects two vertices in both directions.
 
The objective of the minimax path problem is to find a path \( P \) from a starting node \( i \) to a destination node \( j \) such that the maximum weight of the edges in the path \( P \) is minimized. A minimax path distance between a pair of points is the maximum weight in a minimax path between the points (Equation \ref{equ:mmj_defi2}). 

\begin{equation}
\Phi =  \{max\_weight(p) ~ | ~ p \in \Theta_{(i,j,G)}\}
\label{equ:mmj_defi1}
\end{equation}

\begin{equation}
M(i,j~|~G) = min(\Phi)  
\label{equ:mmj_defi2}
\end{equation}
where $ G $ is the undirected dense graph. $\Theta_{(i,j,G)}$ is the set of all paths from node $ i $ to node $ j $.  $ p $ is a path from node $ i $ to node $ j $, $max\_weight(p)$  is the maximum weight in path $p$. $ \Phi $ is the set of all maximum weights. ~ $min(\Phi) $ is the minimum of Set $ \Phi $ \cite{liu2023min}.

The distance can also be called the longest-leg path distance (LLPD) \cite{Little2020} or Min-Max-Jump distance (MMJ distance) \cite{liu2023min}. The all pairs minimax path distances calculate the distance between each pair of points in a dataset $ X $  or graph $ G $ . It is also called all points path distance (APPD) \cite{Little2020}. It is a matrix of shape $ n \times n $. A dataset $ X $ can be straightforwardly converted to a complete graph.

We can use a modified version of the  Floyd–Warshall algorithm to solve the APPD in both directed and undirected dense graphs \cite{weisstein2008floyd},  or use the Algorithm 1 (MMJ distance by recursion) in \cite{liu2023min}, both of them  take $ O(n^3) $ time. However, in an undirected dense graph, we have a better choice. We may use an $ O(n^2) $ algorithm to calculate the APPD matrix. There are several theoretical outcomes which claim the APPD matrix can be solved accurately in $ O(n^2) $  \cite{sibson1973slink,demaine2009cartesian,demaine2014cartesian,alon2024optimal}. However, there is no code implementation of these algorithms, which implies they are impractical.
 
Code implementation is the process of translating a design or algorithm into a programming language. It is critical in algorithm design where ideas are turned into practical, executable code that performs specific tasks.

In section 4.3 (MMJ distance by calculation and copy) of \cite{liu2023min}, the author proposes an algorithm which also claims to solve the APPD matrix accurately in $ O(n^2) $, in an undirected dense graph. The algorithm is referred to as Algorithm 4 (MMJ distance by Calculation and Copy). In the paper, the algorithm is left unimplemented and untested. In this paper, we introduce a code implementation of Algorithm 4, and test it.

The widest path problem is a closely related topic to minimax path problem. In contrary, The objective of the widest path problem is to find a path \( P \) from a starting node \( s \) to a destination node \( t \) such that the minimum weight of the edges in the path \( P \) is maximized. Any algorithm for the widest path problem can be easily transformed into an algorithm for solving the minimax path problem, or vice versa, by reversing the sense of all the weight comparisons performed by the algorithm. Therefore, we can roughly say that the widest path problem and the minimax path problem are equivalent.

 	\begin{figure*} 
	\begin{subfigure}{0.45\textwidth}
	\includegraphics[width=\linewidth]{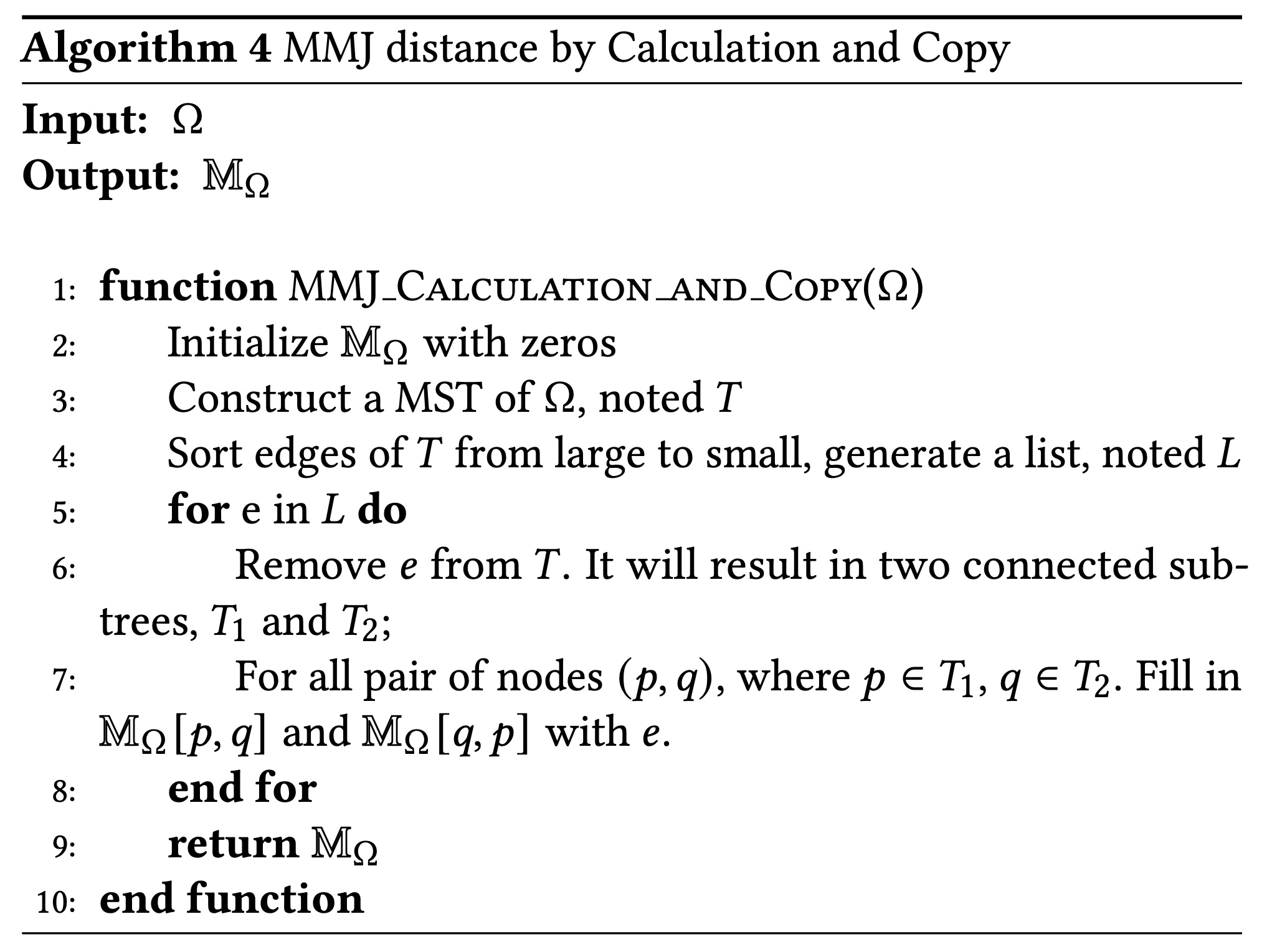}
	\caption{Algorithm 4}   \label{fig:41}
\end{subfigure}    
	\begin{subfigure}{0.45\textwidth}
		\includegraphics[width=\linewidth]{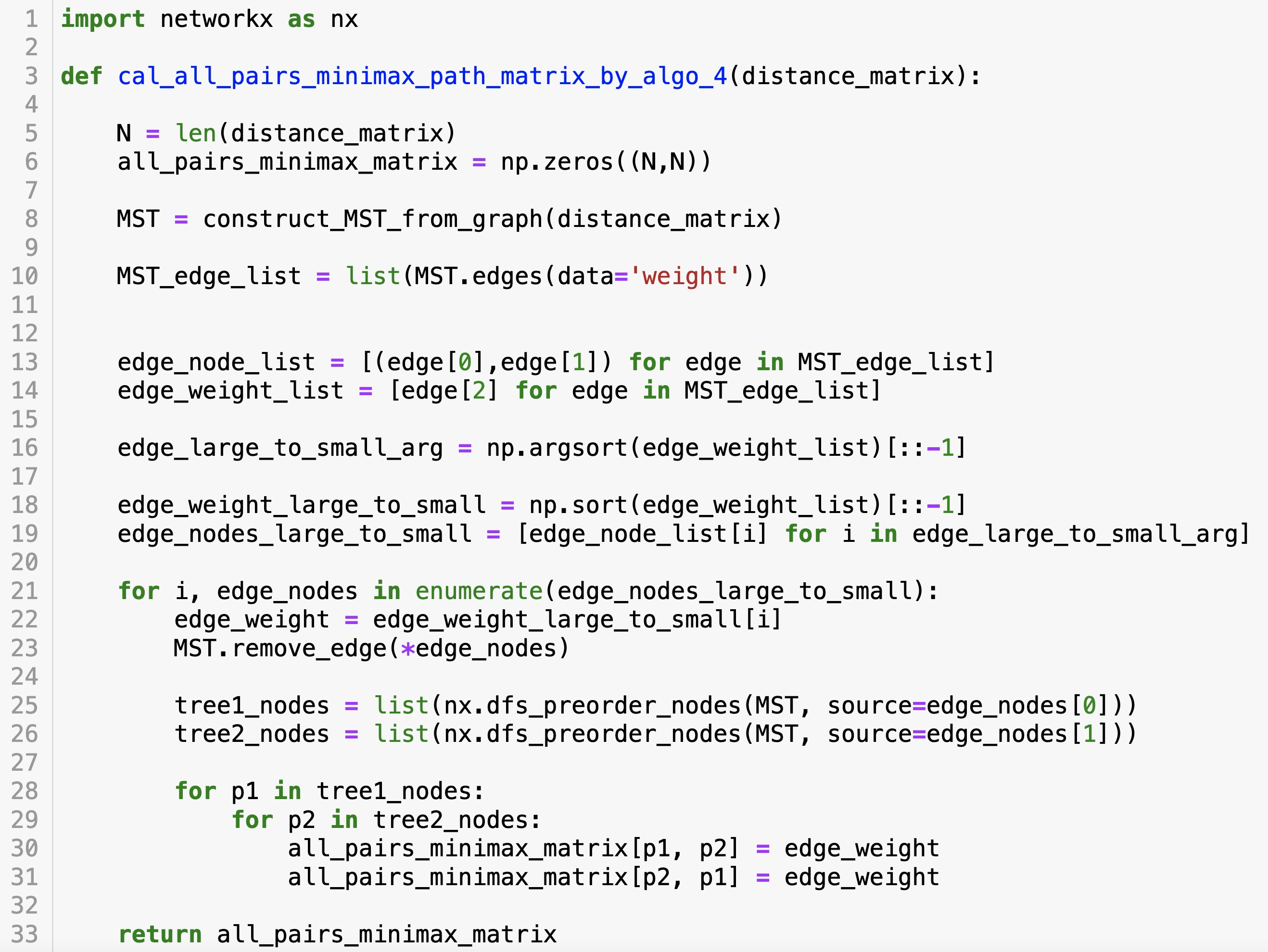}
		\caption{Python implementation of Algorithm 4}   \label{fig:42}
	\end{subfigure}    
 
	\caption{Algorithm 4 and its Python implementation. The three embedded for-loops make it look like an $ O(n^3) $ algorithm, but it is actually an $ O(n^2) $ algorithm.} \label{fig:compare}
\end{figure*}

 	\begin{figure} 
 	\includegraphics[width=0.98\linewidth]{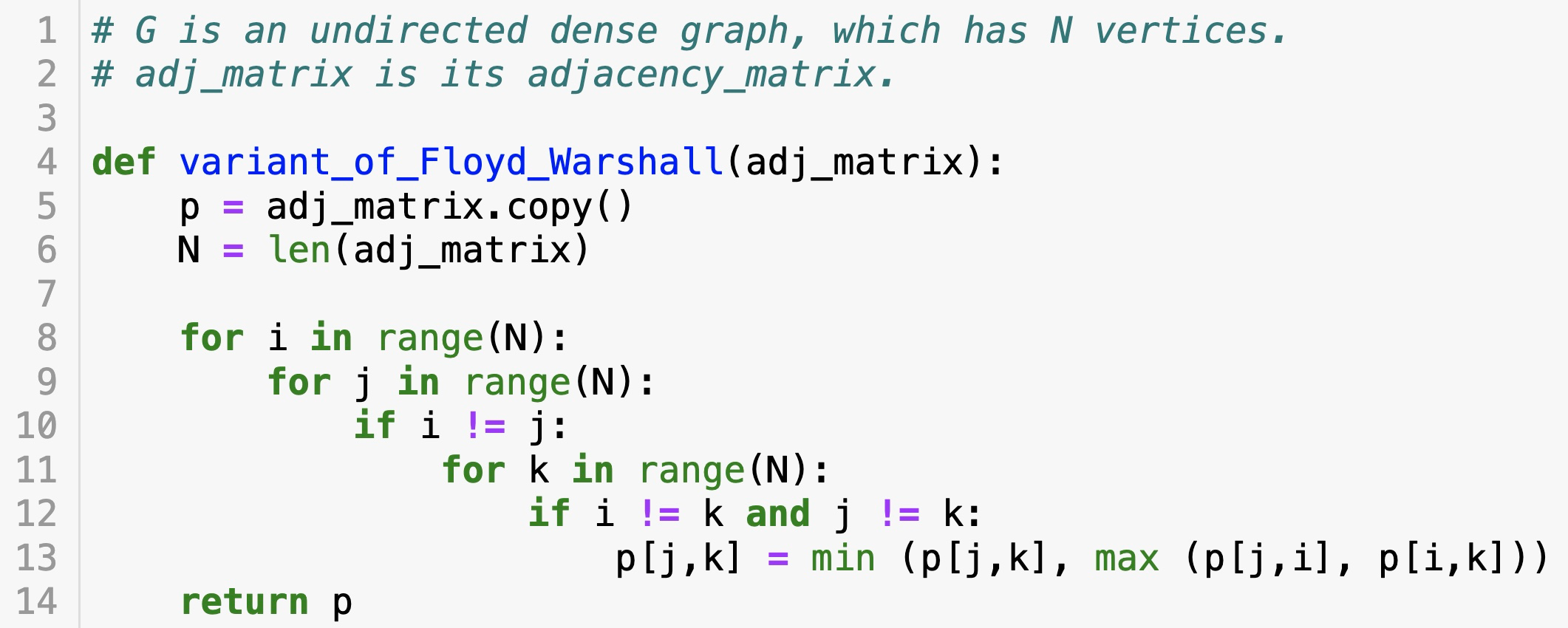}
 	\caption{A variant of the Floyd-Warshall algorithm for solving the minimax path problem} \label{fig:floyd}
 \end{figure}
 
 	\begin{figure} 
 	\includegraphics[width=0.98\linewidth]{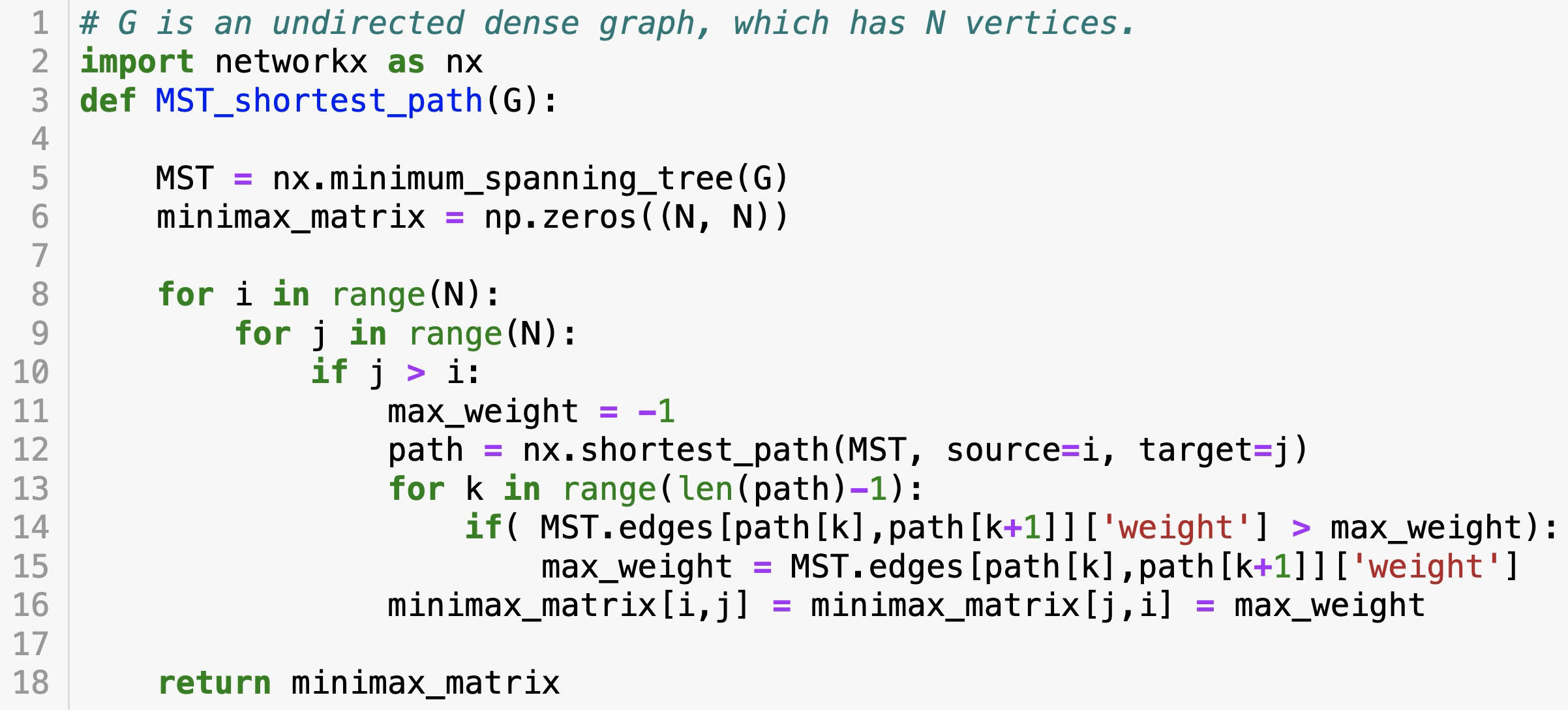}
 	\caption{Python implementation of \textit{MST\_shortest\_path}, see Table \ref{tab:profile}} \label{fig:mst_short}
 \end{figure}

\section{RELATED WORK} \label{Sec_related}

Numerous distance measures have been proposed in the literature, including Euclidean distance, Manhattan Distance, Chebyshev Distance, Minkowski Distance, Hamming Distance,  and cosine similarity. These measures are frequently used in algorithms like k-NN, UMAP, and HDBSCAN. Euclidean distance is the most commonly used metric, while cosine similarity is often employed to address Euclidean distance's issues in high-dimensional spaces.  Although Euclidean distance is widely used and universal, it does not adapt to the geometry of the data, as it is data-independent. Consequently, various data-dependent metrics have been developed, such as diffusion distances \cite{coifman2006diffusion,coifman2005geometric}, which arise from diffusion processes within a dataset, and path-based distances \cite{fischer2003path,chang2008robust}.

Minimax path distance has been used in various machine learning models, such as unsupervised clustering analysis \cite{Little2020,fischer2001path,fischer2003clustering,fischer2003path}, and supervised classification \cite{chehreghani2017classification,liu2023min}.  The distance typically performs well with non-convex and highly elongated clusters, even when noise is present \cite{Little2020}. 

\subsection{Calculation of minimax path distance} 
 The challenge of computing the minimax path distance is known by several names in the literature, such as the maximum capacity path problem, the widest path problem,  the bottleneck edge query problem \cite{pollack1960maximum,hu1961maximum,camerini1978min,gabow1988algorithms}, the longest-leg path distance (LLPD) \cite{Little2020},  and the Min-Max-Jump distance (MMJ distance) \cite{liu2023min}.
 
 A straightforward computation of minimax path distance is computationally expensive due to the large search space \cite{Little2020}. However, for a fixed pair of points \( x \) and \( y \) connected in a graph \( G = G(V, E) \), the distance can be calculated in \( O(|E|) \) time \cite{punnen1991linear}.
 
 A well-known fact about minimax path distance is: “the path between any two nodes in a minimum spanning tree (MST) is a minimax path.”\cite{hu1961maximum} With this conclusion, we can simplify an undirected dense graph into a minimum spanning tree, when calculating the minimax path distance.

 \subsection{Computing the all points path distance} 
 Computing minimax path distance for all points is known as the all points path distance (APPD) problem. Applying the bottleneck spanning tree construction to each point results in an APPD runtime of \( O(\min\{n^2 \log(n) + n|E|, n|E| \log(n)\}) \) \cite{Little2020,camerini1978min,gabow1988algorithms}. The resulting APPD may not be accurate when calculating with bottleneck spanning tree, because a MST (minimum spanning tree) is necessarily a MBST (minimum bottleneck spanning tree), but a MBST is not necessarily a MST. A variant of the Floyd-Warshall algorithm can calculate the APPD accurately in $ O(n^3) $ \cite{aho1974design}.
 Several theoretical results suggest that the APPD matrix can be accurately solved in \( O(n^2) \) time \cite{sibson1973slink, demaine2009cartesian, demaine2014cartesian, alon2024optimal}. However, the absence of code implementations for these algorithms indicates their impracticality.
 
	\begin{figure} 
		\includegraphics[width=0.98\linewidth]{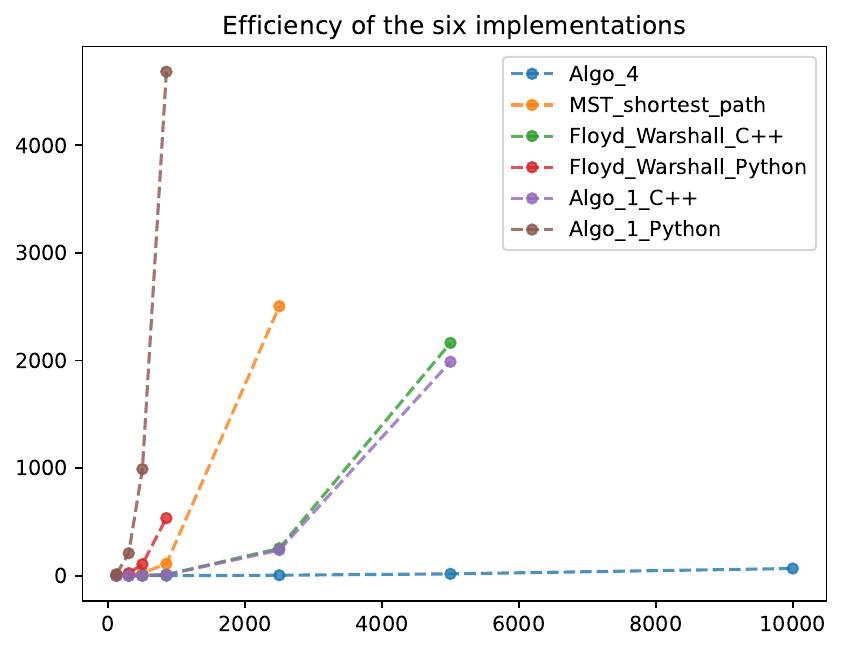}
	\caption{Performance of the algorithms (implementations)} \label{fig:sixsix}
\end{figure}

\begin{table*}
	\begin{center}
		\scalebox{0.8}{
			
\begin{tabular}{lllll}
	\hline
	Implementation ID & Implementation name     & Complexity              & Coding language &  Notes\\ \hline
	0                 & Algo\_1\_Python         & $ O(n^3) $   & Python        &  Algorithm 1 (MMJ distance by recursion) \\
	1                 & Algo\_1\_C++            & $ O(n^3) $   & C++        &    Algorithm 1 (MMJ distance by recursion)  \\
	2                 & Floyd\_Warshall\_Python & $ O(n^3) $   & Python     & A variant of   Floyd-Warshall  Algorithm \\
	3                 & Floyd\_Warshall\_C++    & $ O(n^3) $   & C++       &    A variant of   Floyd-Warshall  Algorithm    \\
	4                 & MST\_shortest\_path          & $ O(n^3log(n)) $   & Python    &  Calculate the shortest path in a MST     \\
	5                 & Algo\_4                 &  $ O(n^2) $  & Python   &  Algorithm 4 (MMJ distance by Calculation and Copy  )     \\ \hline
\end{tabular}		
			
		}		
		
		\caption{Profiles of the four algorithms. Two of them are implemented with different programming languages, Python and C++}
		\label{tab:profile}
	\end{center}
\end{table*}
\begin{table*}
	\begin{center}
		\scalebox{0.7}{
			
 	\begin{tabular}{llllllll}
 	\hline
 	& data 139 (N = 120) & data 109 (N = 300) & data 18 (N = 500) & data 19 (N = 850) & data 16 (N = 2500)  & data 35 (N = 5000)  & data 136 (N = 10000) \\ \hline
 	Algo\_1\_Python         & 13.451s            & 208.363s           & 990.308s          & 4681.911s         & \textgreater{}7200s & \textgreater{}7200s & \textgreater{}7200s  \\
 	Algo\_1\_C++            & 0.033s             & 0.414s             & 1.794s            & 9.032s            & 237.961s            & 1986.928s           & \textgreater{}7200s  \\
 	Floyd\_Warshall\_Python & 1.489s             & 23.353s            & 106.745s          & 534.683s          & \textgreater{}7200s & \textgreater{}7200s & \textgreater{}7200s  \\
 	Floyd\_Warshall\_C++    & 0.033s             & 0.436s             & 2.324s            & 10.035s           & 253.909s            & 2162.514s           & \textgreater{}7200s  \\
 	MST\_shortest\_path          & 0.399s             & 4.229s             & 24.926s           & 110.449s          & 2503.483s           & \textgreater{}7200s & \textgreater{}7200s  \\
 	Algo\_4                 & 0.02s              & 0.073s             & 0.191s            & 0.511s            & 4.311s              & 17.015s             & 67.048s              \\ \hline
 \end{tabular}		

}		
		
		\caption{Performance of the four algorithms. N is the number of points in the datasets.}
		\label{tab:performance}
	\end{center}
\end{table*}

\section{Implementation of the  algorithm}
As described in Section \ref{sec_one}, the Algorithm 4 (MMJ distance by Calculation and Copy) in \cite{liu2023min} also claims to solve the APPD matrix accurately in $ O(n^2) $, in an undirected dense graph. But it is left unimplemented and untested. Figure \ref{fig:41} is Algorithm 4 (MMJ distance by Calculation and Copy) in \cite{liu2023min}, for convenience of reading, we re-post it here. Figure \ref{fig:42} is its python implementation. 

Note the three embedded for-loops make it look like an $ O(n^3) $ algorithm, but it is actually an 
$ O(n^2) $ algorithm. Because when the variable $ i $ in $ Line~ 21 $ is small, both \textit{tree1} and  \textit{tree2} are of size $ O(n) $; but when the variable $ i $ is large, both \textit{tree1} and  \textit{tree2} are of size $ O(1) $. The final net effect is that the three embedded for-loops only access each cell of the APPD matrix only once. Therefore, it is an $ O(n^2) $ algorithm.

In the implementation, we first construct a minimum spanning tree (MST) of the undirected dense graph. The complexity of constructing a MST with prim's algorithm is $ O(n^2) $. Then, we sort the edges of the MST in descending order. It is critical to remove the edges from the MST one-by-one, from large to small. Only by this we can get the two sub-trees, \textit{tree1} and  \textit{tree2}. By traversing each sub-tree, nodes of the two sub-trees can be obtained, respectively.

A further exploration shows we can also compute the APPD matrix by first sorting the edges in ascending order (from small to large), then adding the edges one-by-one to an empty MST tree. \footnote{See the ``Sorting the edges of the MST in ascending order.ipynb" file in \url{https://github.com/mike-liuliu/Algorithm_4}}

\section{Testing of the  algorithm}

In an experiment, we tested the Algorithm 4 (MMJ distance by Calculation and Copy)  on seven datasets with different number of data points, note a dataset can be easily converted to a complete graph. The performance of Algorithm 4 is compared with three other algorithms that can calculate the APPD matrix. 

Table \ref{tab:profile} lists the profiles of the four algorithms. \textit{Algo\_1} is the Algorithm 1 (MMJ distance by recursion) in \cite{liu2023min}, it has complexity of $ O(n^3) $; \textit{Floyd\_Warshall} is a variant of the Floyd-Warshall algorithm. Figure \ref{fig:floyd} is its python implementation. It has complexity of $ O(n^3) $; \textit{MST\_shortest\_path} firstly construct a minimum spanning tree (MST) of the undirected dense graph, then calculate the shortest path between each pair of nodes, then compute the maximum weight on the shortest path. Its complexity is $ O(n^3log(n)) $. Figure \ref{fig:mst_short} is its python implementation. The implementation is based on Madhav-99's code \footnote{\url{https://github.com/Madhav-99/Minimax-Distance}}; \textit{Algo\_4} is Algorithm 4 (MMJ distance by Calculation and Copy) in \cite{liu2023min}, it has complexity of $ O(n^2) $. Both \textit{Algo\_1} and \textit{Floyd\_Warshall} are implemented with C++ and python, respectively, to test the difference between different programming languages.

\subsection{Performance}

Table \ref{tab:performance} is performance of the algorithms (implementations). We test each algorithm with seven datasets which have different number of data points. The data sources corresponding to the data IDs can be found at this URL. \footnote{\url{https://github.com/mike-liuliu/Min-Max-Jump-distance}} The values are the time of calculating the minimax path APPD by each algorithm, on a desktop computer with ``3.3 GHz Quad-Core Intel Core i5'' CPU and 16 GB RAM.

To save time, we stop the execution of an algorithm if  it cannot obtain the APPD matrix in 7200s (two hours). The computing time is recorded only once for each dataset and algorithm. Figure \ref{fig:sixsix} converts the values in Table \ref{tab:performance} into a figure. It can be seen that Algorithm 4 has achieved a good performance than other algorithms. It can calculate the APPD matrix of 10,000 points in about 67 seconds, while other algorithms cannot finish it in two hours.

Reasonably, the C++ implementations of \textit{Algo\_1} and \textit{Floyd\_Warshall} are much faster than their python edition. Interestingly, when implemented in python, \textit{Algo\_1} is much slower than \textit{Floyd\_Warshall}, but a little faster than \textit{Floyd\_Warshall} in C++.
  
\subsection{Solving the widest path problem}
As stated in Section 7 (Solving the widest path problem) of \cite{liu2023min}, Algorithm 4 (MMJ distance by Calculation and Copy) can be revised to solve the widest path problem  APPD in undirected graphs, by constructing a maximum spanning tree and sort the edges in ascending order. In another experiment, we tested using Algorithm 4 to compute the widest path APPD. Result shows Algorithm 4 works good for solving the widest path problem.

\section{Proof of the  algorithm}
A good question is why Algorithm 4 (MMJ distance by Calculation and Copy)  works. Here is a theoretical proof of the correctness of  the algorithm.

Whenever we are about to remove an edge $e$ from the MST, $e$ must belong to a connected sub-tree of MST $T$. The sub-tree is noted $S_t$. A sub-tree is a tree wholly contained in another. Note the MST $T$ can be considered as a sub-tree of itself. We can conclude edge $e$ is the largest edge in sub-tree $S_t$. Since the edges have been sorted in descending order, and edges larger than $e$ have been removed in previous steps. It does not matter if there are other edges in $S_t$ which are as large as $e$.
 
After removing edge $e$ from $S_t$, we get two smaller connected sub-trees, $tree1$ and $tree2$. For any pair of nodes $ (p,q) $, where $p \in  tree1$, $q \in  tree2$, the minimax path distance between $p$ and $q$ must be the weight of edge $e$. Because “the path between any two nodes in a minimum spanning tree (MST) is a minimax path” \cite{hu1961maximum}, and edge $e$ is the largest edge in sub-tree $S_t$. A path between $p$ and $q$ must pass through edge $e$, and edge $e$ is the largest edge in the path. It does not matter if there are other edges in the path which are as large as $e$. Note a sub-tree that has only one node is considered as a valid sub-tree.

Therefore, the minimax path distance between $p$ and $q$ must be the weight of edge $e$. The correctness of  Algorithm 4 (MMJ distance by Calculation and Copy) is proved.

\section{Using parallel programming}
If speed is the main concern of calculating the APPD matrix, we can use parallel programming to accelerate Algorithm 4. Firstly, we can use different processors for traversing the \textit{tree1} and  \textit{tree2}  in $ Line~ 25~  and ~ 26 $ of Figure \ref{fig:42}. Secondly, we can copy the minimum spanning tree (MST) to many processors. For the $ n $th processor, we just remove the $ n $ largest edges, obtaining the $ n $th \textit{tree1} and  \textit{tree2}, traversing them, then fill in the corresponding positions of  the APPD matrix that are decided by the $ n $th \textit{tree1} and  \textit{tree2}.

\subsection{Description of variants of Algorithm 13}
The parallel accelerated version of Algorithm 4 (MMJ distance by Calculation and Copy), is called Algorithm 13 (APPD accelerated by parallel computing). It distributes the computation of the MMJ matrix across multiple threads.

In Variant 1 of Algorithm 13, the task queue initially stores all minimum spanning tree (MST) edge indices in descending order of edge weights. The main thread continuously retrieves tasks from the front of the queue, while worker threads retrieve tasks from the back, enabling simultaneous processing from both directions. The main thread starts with a fully active MST and processes each task by temporarily removing the corresponding edge, computing the two resulting connected components (tree1 and tree2) using DFS, and filling the corresponding entries of the MMJ matrix with the removed edge weight. In contrast, each worker thread starts with an empty active MST and incrementally activates edges according to the task ID it receives, thereby reconstructing the required partial MST by adding edges one by one. After activating the necessary edges, the worker computes the node sets of tree1 and tree2 via DFS and updates all pairwise entries between the two node sets in the MMJ matrix using the associated edge weight.

Variant 2 of Algorithm 13 is a fully worker-driven parallel implementation of Algorithm 4 (MMJ distance by Calculation and Copy), in which all MMJ matrix computations are performed by worker threads, while the main thread only initializes the data structures and launches the workers. The task queue initially stores all minimum spanning tree (MST) edge indices in descending order of edge weights, and each worker thread continuously retrieves tasks from the front of the queue. Unlike Variant 1, every worker thread starts with a fully active MST and progressively removes edges according to the task IDs it receives. After removing the required edges, the worker computes the two connected components (tree1 and tree2) generated by the removed edge using DFS, and then updates all pairwise entries between the two node sets in the MMJ matrix using the associated edge weight. Other variants of Algorithm 13 follow similar parallel processing principles with different task allocation and MST update strategies.

Variant 4 of Algorithm 13 is a fully worker-driven parallel implementation of Algorithm 4 (MMJ distance by Calculation and Copy) based on incremental MST construction. The MST edges are first sorted in ascending order of edge weights, and the task queue initially stores their indices from the smallest to the largest. Each worker thread starts with an empty local MST and continuously retrieves tasks from the front of the shared queue. For a task associated with the k-th MST edge, the worker incrementally adds all preceding edges, from the smallest-weight edge through the (k-1)-th edge, to its local MST, while leaving the current edge uninserted. The two endpoints of the current edge therefore belong to two separate connected components. The worker identifies these components, denoted by tree1 and tree2, using DFS and then assigns the current edge weight to all corresponding pairwise entries between the two node sets in the MMJ matrix. Because each worker preserves its partially constructed local MST between successive tasks and the task indices obtained by a worker increase monotonically, only newly required edges need to be added. All MMJ matrix calculations are performed by the worker threads, while the main thread initializes the data structures, constructs the task queue, launches the workers, and waits for their completion.

Variant 5 is closely related to Variant 4. For each MST edge processed in nondecreasing order of weight, Variant 4 explicitly constructs the forest containing all preceding edges and performs two graph traversals to identify the components connected by the current edge. It then assigns the current edge weight to all cross-component vertex pairs. Variant 5 precomputes the same sequence of component merges as a Kruskal reconstruction tree. Each internal reconstruction-tree node corresponds exactly to one edge-processing task in Variant 4, and its two child subtrees correspond to the two components identified by the traversals in Variant 4. By arranging the original vertices in reconstruction-tree leaf order, these components become contiguous intervals, allowing the corresponding matrix regions to be filled using sequential memory operations. Threads own disjoint output-row ranges, eliminating the dynamic task queue and preventing concurrent writes to the same output row. Here\footnote{\url{https://github.com/mike-liuliu/Algorithm_4/tree/main/Algorithm_4_in_cpp_parallel_computing}} are C++ code implementations of variants of Algorithm 13.

\section{Discussion} 
\subsection{Merit of Algorithm 1}
Algorithm 1 (MMJ distance by recursion) has a merit of warm-start. Suppose we have calculated the APPD matrix $ M_G $ of a large graph $ G $, then we got a new point (or node) $ p $,  where $ p \notin G$. The new graph is noted $G + p$. To calculate the APPD matrix of graph $G + p$, if we use other algorithms, we may need to start from zero. Algorithm 1 has the merit of utilizing the calculated $ M_G $ for computing the new APPD matrix, with the conclusions of Theorem 3.3., 3.5., 6.1., and Corollary 3.4. in \cite{liu2023min}. This is especially useful when the graph is a directed dense graph, where starting from zero needs $ O(n^3) $ complexity, but a warm-start of Algorithm 1 (MMJ distance by recursion) only needs $ O(n^2) $ complexity. We can say Algorithm 1 supports online machine learning\footnote{\url{https://en.wikipedia.org/wiki/Online_machine_learning}}, in which data becomes available in a sequential order.

\section{Conclusion} 
We implemented the Algorithm 4 (MMJ distance by Calculation and Copy)  that was introduced in a previous paper. Then tested the implementation and compared it with several other algorithms that can calculate the all pairs minimax path distances, or also called the all points path distance (APPD). Experiment shows Algorithm 4 works good for solving the widest path or minimax path APPD matrix. As an algorithm of $ O(n^2) $ complexity, it can drastically improve the efficiency of calculating the APPD matrix. Note algorithms for solving the APPD matrix are at least in $ O(n^2) $ complexity, because the matrix is an $ n \times n $  matrix.

In Section 2.3.3. of the paper  ``Path-Based Spectral Clustering: Guarantees, Robustness to Outliers, and Fast Algorithms," \cite{Little2020} Dr. Murphy and his collaborators write:

\textit{``Naively applying the bottleneck spanning tree construction to each point gives an APPD runtime of $ O(min\{n^2 log(n)+n|E|, n|E| log(n)\}) $. However the APPD distance matrix can be computed in $ O(n^2) $, for example with a modified SLINK algorithm (Sibson, 1973), or with Cartesian trees (Alon and Schieber, 1987; Demaine et al., 2009, 2014). "}

The author sent an email for further clarity about this statement. 

The author:

\textit{``You indicated the APPD distance matrix can be computed in $ O(n^2) $. However, I searched the Internet and github, I have not found any code implementation that can accurately calculate the APPD distance matrix in $ O(n^2) $. Do you know any code implementation of that? Please indicate it to me. "}

Dr. Murphy:

\textit{``If you can find an implementation of SLINK to do single linkage clustering in $ O(n^2) $, then you can do APPD by reading off the distances from the resulting dendrogram.  I don't know any implementations of SLINK, and it may be easier to prove things about than to implement practically. "}

\textit{``Regarding tree structures, these are certainly more of theoretical interest, and I would not be surprised if there were no practical implementations of them at all.  So, achieving $ O(n^2) $ via those methods may be impractical. "}

Therefore, we can roughly conclude:  the Algorithm 4 (MMJ distance by Calculation and Copy) in \cite{liu2023min}, is the first algorithm that has actual code implementation to solve the APPD matrix of minimax path or widest path problem in $ O(n^2) $, in an undirected dense graph.
 
\bibliographystyle{ACM-Reference-Format}
\bibliography{mmj} 
 
\end{document}